\pdfoutput=1
\documentclass{iau}

\title[Proceedings of the International Astronomical Union]
{High-Mass X-ray Binaries: progenitors of double compact objects} 

\author[Edward P.J. van den Heuvel]   
{Edward P.J. van den Heuvel}

\affiliation{Anton Pannekoek Institute of Astronomy, University of Amsterdam, \\ Postbus 92429,
NL-1090GE, Amsterdam, the Netherlands \\ email: {\tt E.P.J.vandenHeuvel@uva.nl} \\[\affilskip]}

\pubyear{2019}
\volume{346}  
\jname{High-Mass X-ray Binaries}
\editors{L.M. Oskinova, E. Bozzo,  T. Bulik,  D. Gies, eds.}
\usepackage{natbib}
\usepackage{graphicx}

\begin{document}

\maketitle

\begin{abstract}
A summary is given of the present state of our knowledge of High-Mass X-ray Binaries (HMXBs), their formation and expected future evolution. Among the HMXB-systems that contain neutron stars, only 
those that have orbital periods upwards of one year will survive the Common-Envelope (CE) evolution that follows the HMXB phase. These systems may produce close double neutron stars with eccentric 
orbits. The HMXBs that contain black holes do not necessarily evolve into a CE phase.  Systems with relatively short orbital periods will evolve by stable Roche-lobe overflow to short-period 
Wolf-Rayet (WR) X-ray binaries containing a black hole. Two other ways for the formation of WR X-ray binaries with black holes are identified: CE-evolution of wide HMXBs and homogeneous evolution of 
very close systems. In all three cases, the final product of the WR X-ray binary will be a double black hole or a black hole – neutron star binary.     

\end{abstract}

\section{Introduction}

My emphasis in this review is on evolution: on what we think to know about how High Mass X-ray Binaries (HMXBs) were formed and how they may evolve further to form binaries consisting of two compact 
objects: double neutron stars, double black holes and neutron star-black hole binaries. After a brief description in section 2 of the different types of HMXBs, I summarize in section 3 what I 
consider 
to be the most important new developments in the field of HMXBs of the past one-and-a-half decades. In section 4, I describe the past evolution of HMXBs and give a rough outline of the expected 
future evolution of HMXBs that contain neutron stars, as we  presently think we understand them.

In the later evolution of the HMXBs that contain neutron stars, Common-Envelope Evolution (CEE) is expected to play a crucial role. It is expected that only wide neutron-star HMXBs, with orbital 
periods upwards of about one year, will survive as binaries, with very short orbital periods and consisting of a helium star and a neutron star. Such systems later evolve into close eccentric-orbit 
double neutron stars, of which presently some twenty are known.

Neutron-star HMXBs with orbital periods shorter than about one year will merge into a single object, possibly resembling a Thorne-Zytkow star. In section 5 the predictions of the “standard model” for 
the formation and further evolution of neutron-star HMXBs are compared with the observations. 

Section 6 summarizes our knowledge of the black-hole X-ray binaries and Wolf-Rayet X-ray Binaries (WRXBs), and focuses on the different ways in which BH-HMXBs may evolve into WRXBs, and on possible 
other channels for the formation of WRXBs. In HMXBs with short orbital periods that contain a black hole, stable Roche-lobe overflow is possible, such that CEE can be avoided, and the systems may 
survive as close binaries consisting of a helium star (WR star) and a black hole. It is argued that the compact stars in practically all WRXBs must be black holes, making these systems ideal 
progenitor systems of double black holes and Black hole-Neutron star (BH-NS) binaries.  

\section {The different types of High-Mass X-ray Binaries}

There are three main types of HMXBs,  with the following characteristics:

(i)	The first type, discovered by \citet[]{Schreier1972} and \citet[]{WebsterMurdin72}, is that of the supergiant HMXBs. In these systems the donor 
star is an O- or early B-type supergiant star that is close to filling its Roche lobe. The orbital periods of these systems, many of which are eclipsing, 
are mostly shorter than 15 days. These systems 
are persistent (permanent) X-ray sources, mostly powered by the capture of matter from the strong stellar wind of the supergiant companion. In a few 
cases, such as the $2.1$\,d orbit eclipsing and 
regularly pulsating X-ray source Centaurus X-3 \citep[]{Schreier1972}, the X-ray source is powered by beginning Roche-lobe overflow. 
The supergiant HMXBs are relatively rare, 
their total known number in the Galaxy being about 30, and, in practically all of them, the compact star is a neutron star: a X-ray pulsar. The blue supergiants
have masses typically in the range $20$ 
to $50 {M_{\rm\odot}}$. 

(ii)	 The second type of systems, practically all containing neutron stars, is that of the B-emission X-ray Binaries (short: BeXBs), discovered in 1975 with the Ariel V satellite, and first 
recognized and explained as a separate class by \citet[]{Maraschi1976}. Most of these systems are recurrent transients, which can be quiet for many decades and then suddenly 
flare up as a strong pulsating X-ray source for weeks to months. The companion stars here are rapidly rotating B-stars that are in or very close to the main sequence and are deep inside their Roche 
lobes; they have a variable emission-line spectrum of hydrogen. These lines are formed in a rapidly rotating disk of gas that surrounds the star in its equatorial plane. 
The emission lines may be 
absent for years,  then return, due to ejection of gas from the equatorial regions of the star (see Rivinius, this volume). If the Be-star has a compact 
companion, the motion of the latter through 
this ejected equatorial disk of gas will cause it to accrete matter and temporarily become a strong X-ray source.
The BeXBs tend to have relatively long orbital periods, ranging from about 15 days to over $4$ years. The Be stars in these systems typically
have masses in the range 8 to $20{M_{\rm\odot}}$. 

The BeXBs form the largest group of HMXBs, with a presently recognized number of around $220$. Particularly the SMC is very rich in these systems, with a total number at least $120$ (Haberl, this 
volume). In about half of the BeXBs regular X-ray pulsations have been observed, and the other half have similar 
X-ray spectra, suggesting they  also harbor neutron stars. Only one Black hole BeXB 
is known \citep[][see also Ribo, this volume]{Casares2014}.

(iii)	The third class of HMXBs as that of the Wolf-Rayet X-ray Binaries (WRXBs), of which presently only seven are known. 
Except for one, Cyg X-3, they are all located in external galaxies \citep[][see also Carpano, this volume, and Soria, this volume]{Esposito2015}. With the exception of 
the system M101 ULX-1, they all have very short orbital periods, of around 
one day or less (see Table \ref{tab2}). Wolf-Rayet (WR) stars are helium stars, and the very strong emission lines of He, 
C and N which are characteristic for these stars, are produced in an 
extremely strong radiation-driven stellar wind, with mass-loss rates around $10^{-5}\,M_\odot$\,yr$^{-1}$, 
and velocities $2000-5000$\,km\,s$^{-1}$ \citep[e.g.][]{Hamann2006, Crowther2007, Conti2008}. As the measured masses of 
WR-star in binaries are at least $8 {M_{\rm\odot}}$ \citep[]{Crowther2007, Shenar2016}, the large luminosity and strong radiation 
pressure required for driving WR winds apparently develop only in helium stars with masses above this lower mass limit. 
To produce a helium core larger than $8{M_{\rm\odot}}$, the progenitor star of 
the WR star must have had a mass of at least $30{M_{\rm\odot}}$. WRXBs must therefore have had HMXB progenitors with 
donor masses above $30{M_{\rm\odot}}$. As we will show in section 6, based on 
arguments from binary evolution, the compact stars in WRXBs most likely are black holes \citep{vandenHeuveletal2017}.

\section{Important developments in the HMXB field in the past one-and-a-half decades}
I list here what in my personal view were the most important developments in the HMXB field in the past 15 years:

1.	The discovery with the INTEGRAL satellite of two new classes of supergiant HMXBs, which increased the known galactic number of supergiant systems by a factor of four (see Sidoli, this 
volume): (a) the highly obscured supergiant systems, and (b) the Supergiant Fast X-ray Transients (SFXTs).
These discoveries are an illustration of the fact that astronomy is a science that is heavily affected by observational selection effects. Earlier X-ray survey missions were sensitive for relatively 
soft X-rays, with energies below about 10\,keV, where X-ray absorption by neutral hydrogen plays an important role. Objects that are highly obscured due to a high hydrogen column density towards the 
source, were missed by these missions. The IBIS/ISGRI soft gamma-ray telescope of INTEGRAL is in fact a hard X-ray telescope, working in a spectral region above 15\,keV, where X-ray absorption by 
neutral hydrogen is much less important. IBIS discovered many new supergiant HMXBs with hydrogen column density in or around the system larger than $10^{23}$ H-atoms\,cm$^{-2}$ such that X-rays below 
10\,keV are very heavily absorbed. A key example is the first such source discovered IGR J16318-4848 with ${N_{\rm H}}= 2\times 10^{24}$\,cm$^{-2}$ \citep{Walter2003}.  
Furthermore, with the large field of view ($30 \times 30$\,degree$^{2}$) of IBIS and the long stretches of observing times required for the gamma-ray spectroscopy with the SPI telescope, which looks 
in the 
same direction as the IBIS telescope, IBIS also turned out to be an excellent instrument to detect short-duration hard X-ray flares when these happen. These flares, lasting only a fraction of a day, 
are the special property of the Supergiant Fast X-ray Transients. Earlier instruments never stared for sufficiently long time intervals at the same field on the sky, and therefore missed these 
short-lasting very intense flares, which turned out to arise from this new class of supergiant HMXBs. In the Corbet diagram of High-Mass X-ray Binaries, in which the pulse periods of the supergiant 
systems and BeXBs are plotted against their orbital periods \citep{Sidoli2012}, one finds that the new supergiant systems occupy the same region of the diagram as the 
earlier-discovered supergiant systems, the only difference being that now a few systems with quite long orbital periods have been added.

2.	The discovery of extragalactic Wolf-Rayet X-ray binaries (see section 6 and Carpano, this volume, and Soria, this volume), and of the first Black Hole BeXB \citep[][Ribo, this 
volume]{Casares2014};

3.	The discovery of hundreds of extragalactic Ultra-Luminous X-ray sources (ULXs); I refer to the papers in this volume by Harrison, by Heida and by Walter, and to the many poster presentations 
of this conference;

4.	The discovery of a class of BeXBs with nearly circular orbits \citep{Pfahl2002}, which demonstrates that there is a class of neutron stars that receive hardly any 
velocity kick in their birth events. The same class has now also been recognized among the double neutron stars, a considerable fraction of which have very low orbital eccentricities, indicating that 
the second-born neutron star received hardly any kick at birth \citep[e.g.][]{Tauris2017}. An example is the double pulsar PSR J0737-3039, which has e=0.088. These 
low-kick 
neutron stars may either have been formed by electron-capture collapse, or by the collapse of ultra-stripped iron cores \citep[]{Tauris2015}. Electron-capture collapse may possibly 
occur only in binary systems \citep{Podsiadlowski2004, Dessart2006, Kitaura2006}, and also the formation of 
ultra-stripped cores requires binary interaction.

5.	The development of the “Settling Accretion Theory” for magnetized neutron stars \citep{Shakura2012}. This theory gives a consistent explanation of how neutron stars 
accreting from the stellar wind of a companion can be spun down to very long spin periods. It also can give an explanation for the SFXT outbursts of supergiant HMXBs \citep[][see also Postnov, this 
volume]{Postnov2014}.  

\section{Past and future evolution of HMXBs}
\begin{figure}[b]
\begin{center}
 \includegraphics[width=5.2in]{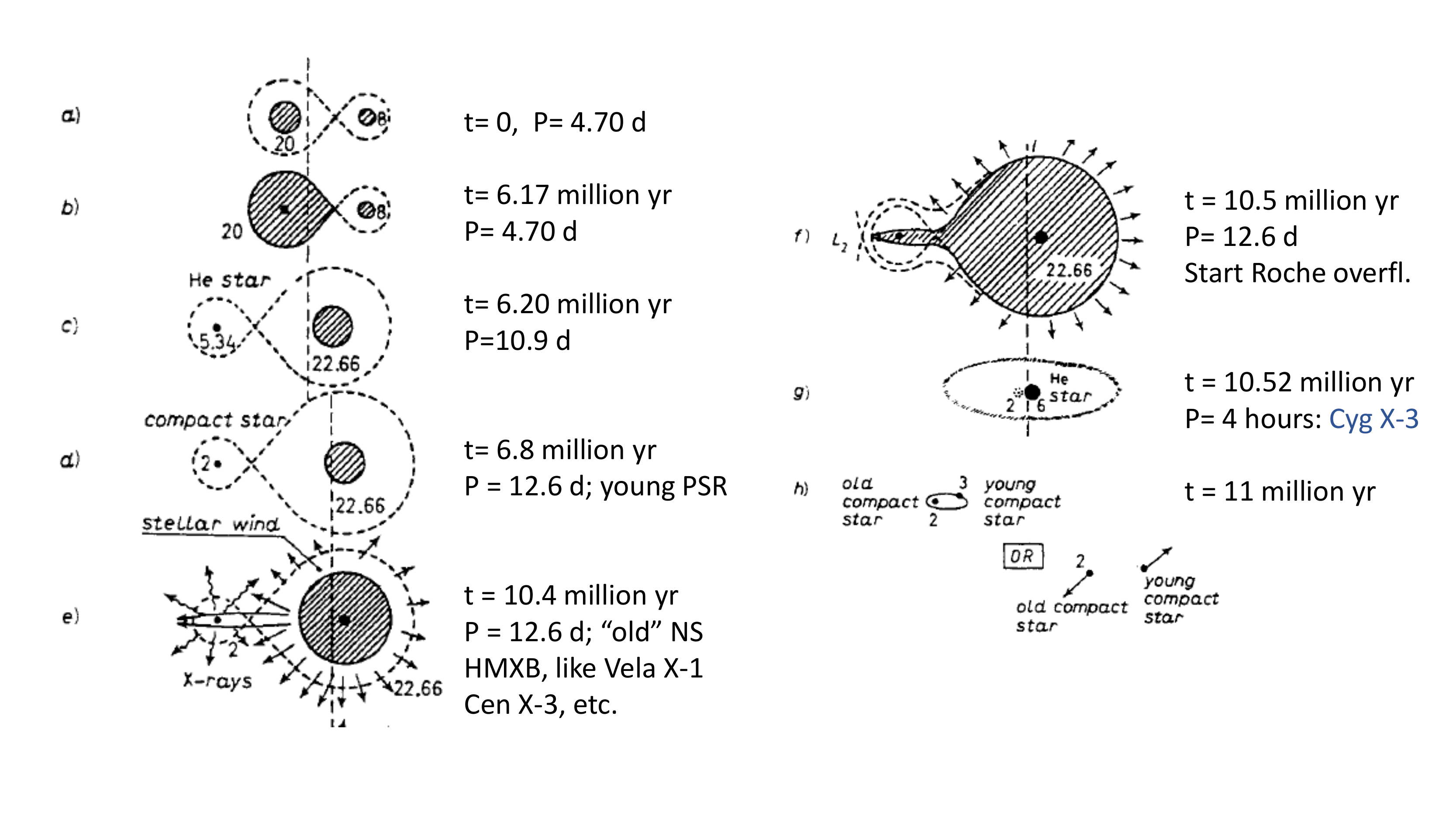} 
 \caption{Model of the evolution of a close binary with initial components of ${20}$ and ${8 {M_{\rm\odot}}}$ into 
 a supergiant HMXB (left) and its further evolution into a close binary consisting of 
two compact stars (right) \citep[after][]{vandenHeuvel1976}. Numbers near the stars indicate stellar masses. 
As explained in the text, in case the first-formed compact star is 
a 
neutron star, survival of stage {\it f} of the evolution requires that the orbital period of the 
system is much longer than indicated here; see also Fig.\,2.}
   \label{fig1}
\end{center}
\end{figure}

The basic model for the formation and later evolution of HMXBs, as depicted in Fig.\,\ref{fig1}  \citep[from][]{vandenHeuvel1976}), was developed in the years 1972-1974 
\citep{HeuvelHeise1972, TutukovYungelson1973, HeuvelLoore1973, FlanneryHeuvel1975, Loore1975}. 

The essential ingredient of the model for the formation of HMXBs is the occurrence of extensive mass transfer in the binary prior to the first supernova explosion in the system, such that by the time 
of this explosion the exploding star has become the less massive component of the binary. The supernova mass ejection (assumed to occur in a spherically symmetric way) will then not disrupt the 
binary, because if less than half of the total mass of the system is explosively ejected, the binary remains bound \citep{Blaauw1961}. This is simply a consequence of the virial 
theorem. In the case that the less massive star of the binary explodes, the orbit will become eccentric and the system becomes a runaway star (“slingshot effect”); the orbital changes and runaway 
velocities for this case were calculated by \citet{Heuvel1968}. If the compact object formed in the supernova explosion receives a kick-velocity at birth, the system may still 
be 
disrupted, even if it is the less massive component that explodes. The first calculations of the effects of birth-kicks on the orbits were made 
by \citet{FlanneryHeuvel1975}.

As to the expected further evolution of a HMXB after the massive companion of the compact star begins to overflow its Roche lobe: \citet{HeuvelLoore1973} had 
assumed, as depicted in Fig.\,1\,f-h, that a HMXBs with a donor star with a radiative envelope and a short orbital period could evolve with stable Roche-lobe overflow.  In this case the entire 
H-rich 
envelope of the donor is lost, carrying off much orbital angular momentum, such that a very narrow system will remain, consisting of a helium star (the helium core of the donor) and the compact star. 
We suggested that the highly peculiar X-ray binary Cygnus X-3, with a 4.8\,h orbital period, is such a system, which was confirmed 19 years later by IR spectroscopy which showed its companion to 
be 
a Wolf-Rayet star of type WN5 \citep{Kerkwijk1992}.  Such a close system consisting of a helium star and a compact star will, after the supernova explosion of the helium 
star - if not disrupted - produce a close eccentric binary consisting of two compact stars. When \citet{HulseTaylor1975} discovered the first double neutron star, with a very 
short orbital period (P = 7h\,45m) and high orbital eccentricity (e=0.615), it therefore was clear to us that this must be a later evolutionary product of a HMXB, after it went in Roche-lobe overflow 
and spiraled-in \citep{FlanneryHeuvel1975,Loore1975}.

Although this picture looked straightforward, it was shown somewhat later by \citet{Paczynski1976} that, because of the extreme mass ratio of a system consisting of a blue supergiant 
and a neutron star, Roche-lobe overflow in such a system is unstable, and once Roche-lobe overflow starts, the mass transfer will run out of hand, and the neutron star will be engulfed by the 
envelope of the massive star, such that a Common Envelope will form, in which the neutron star and the compact helium core of the massive star spiral-in towards each other, due to the large friction 
on their orbital motion in the envelope. Numerical computations of the spiral-in process were pioneered by \citet{TaamBO1978}, and further developed by Taam and 
his collaborators over many years \citep[e.g.][]{TaamSandquist2000} and others. For recent developments in the CEE field I refer to the paper by Ricker in this 
volume. The 
computations by Taam and collaborators showed that, for a neutron-star HMXB to survive Common Envelope Evolution (CEE), the system must start out with an orbital period longer than about one year 
\citep{Taam1996}.
For these reasons, the system of Fig.\,1-f will not survive spiral in: the neutron star will spiral into the core of its companion, leading to a single massive star with a neutron star in its center, 
a so-called Thorne-Zytkow star \citep{ThorneZytkow1977}. Computations of the structure of such stars showed that they are expected to look like red supergiants with 
peculiar 
element abundances, particularly of p-process elements \citep{Podsi1995}.

\begin{figure}[b]
\begin{center}
\includegraphics[width=3.4in]{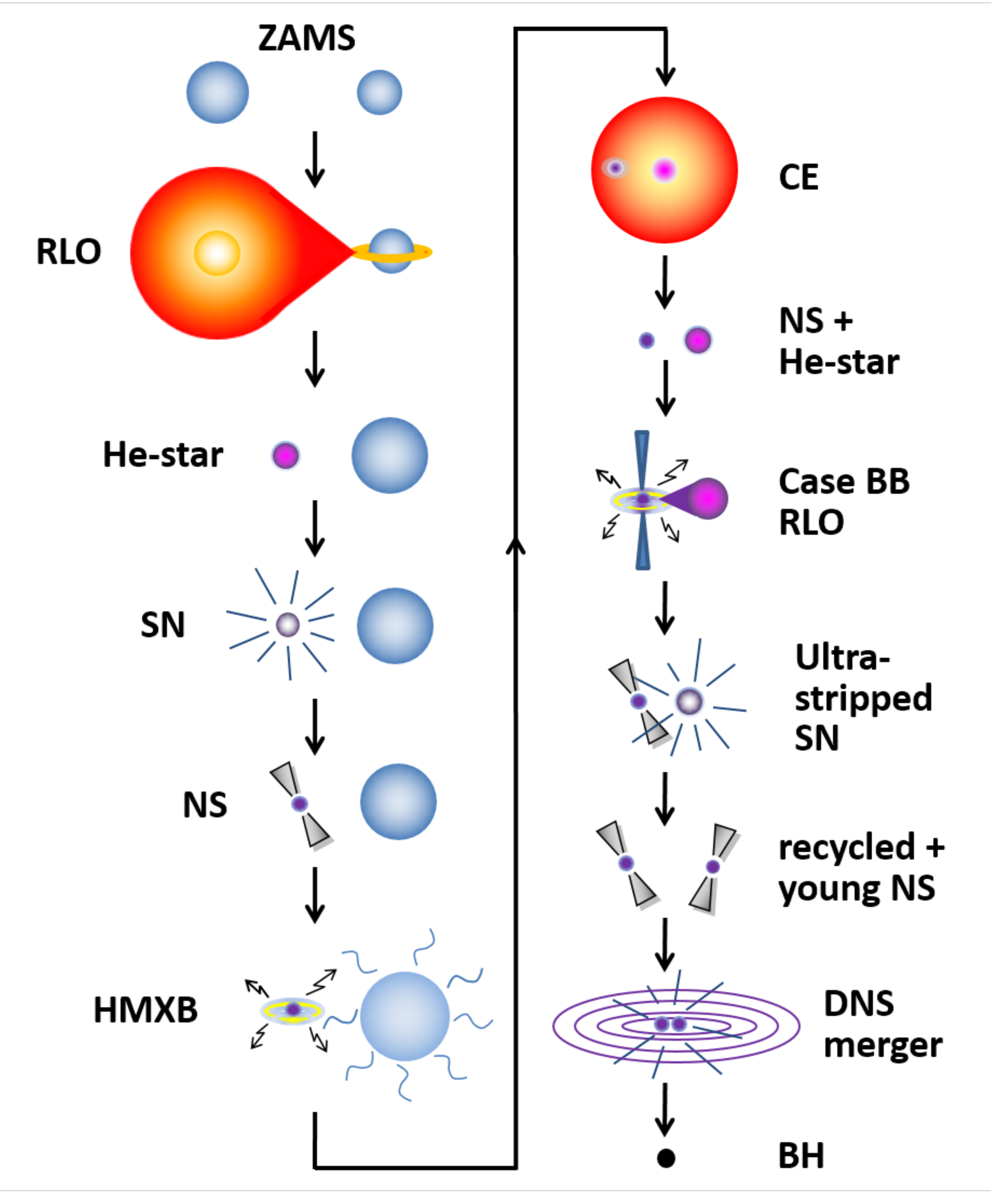} 
 \caption{Evolution of a wide neutron-star HMXB, with orbital period longer than about one year, into a close double neutron star. The wide HMXB evolves through a Common Envelope phase into a close 
helium star plus neutron star binary, which produces a close double neutron star. This system, consisting of an old recycled neutron star and a newborn neutron star, may finally merge into a black 
hole \citep[from][]{Tauris2017}.}
   \label{fig2}
\end{center}
\end{figure}

The only HMXBs with orbital periods longer than about one year are the BeXBs. Therefore, the progenitors of the double neutron stars are the long-period BeXBs. As the supergiant neutron-star HMXBs 
almost all have relatively short orbital periods, they will not survive CEE, and are expected to terminate as single Thorne-Zytkow stars. So far, such stars have never been identified with certainty. 
Still, as the galactic formation rate of HMXBs is of the order of $2\times 10^{-4}$\,yr$^{-1}$ , one would expect Thorne-Zytkow stars to be formed at about the same rate, leading to some 20 to 200 
such stars in 
the Galaxy \citep{Podsi1995}. The absence of evidence of their existence remains a great puzzle.  
Fig.\,\ref{fig2} depicts the generally accepted model for formation of a close double neutron star as a later evolution product of a BeXB \citep{Tauris2017}. Presently, some 20 
of such binaries are known (15 of them were listed in the above-mentioned paper by Tauris et al.). About half of them have orbital periods such that they will merge by GW losses within a Hubble time. 

If the compact star in the evolutionary picture of Fig.\,1f-h is not a neutron star, but a black hole with a mass above 20 to 30 per cent of the mass of the blue supergiant companion, the Roche-lobe 
overflow is stable \citep{vandenHeuveletal2017, Pavlovski2017}, and also systems with a short orbital period will survive as a close system 
consisting of a WR-star (helium star) and a black hole \citep{vandenHeuveletal2017}. So, for black-hole HMXBs the evolutionary picture of Fig\,1f-h is valid. I deal 
with this type of evolution in more detail in section 6.

\section{Predictions of the standard model for formation and evolution of HMXBs, compared to the observations}

The model depicted in Fig\,\ref{fig1} makes the following predictions:

(i)	HMXBs must be runaway stars \citep{HeuvelHeise1972, TutukovYungelson1973};

(ii)	There must be massive stars with a young pulsar companion \citep{Heuvel1974};

(iii)	Since many young neutron stars receive velocity kicks of several hundreds of km\,s$^{-1}$ at birth, there must also be disrupted systems;

(iv)	There must be double neutron stars in which one of the stars is a young strong-magnetic-field pulsar \citep{SriniHeuvel1982}.

All of these predictions have in later years been confirmed by the observations.
The first prediction was confirmed by the work of \citet{Kaper1997} who found the bow shock of the Vela X-1 (4U0900-40) supergiant HMXB, and measured its 45\,km\,s$^{-1}$ excess 
transverse 
velocity, indicating that the system originated in the association Vela OB1 some 2 to 3 million years ago. Later also a high excess transverse velocities were found for two other supergiant systems: 
76\,km\,s$^{-1}$ for 4U1700-37 (its likely origin is in the association Sco OB1), and 85\,km\,s$^{-1}$ for 4U1538-52 \citep{Kaper2001}. For the BeXBs the runaway velocities are much 
smaller, as expected because of their lower masses and wider orbits \citep{Heuvel2001}.

As to the second prediction: the discovery of gamma-ray emission from several OB binaries has shown that indeed there are OB stars that must have young pulsar companions \citep{Dubus2017}. 
Young Crab-like pulsars emit a highly relativistic electron-positron pulsar wind. Both positron annihilation and inverse Compton boosting of optical 
photons from the OB star by the relativistic electrons lead to the emission of gamma rays. Presently, there are 7 such systems known.

Two of the systems (LS 5039 and LMC P3) are short-period O-type binaries that are ideal progenitors of the supergiant HMXBs. The other ones have wide orbits and are ideal progenitors for BeXBs.

The third prediction was beautifully confirmed by the discovery by \citet{Dincel2016} that in the supernova remnant Semeis 147 there is a B0V runaway star with an excess 
transverse velocity of $74(\pm 7.5)$\,km\,s$^{-1}$, plus a high-velocity pulsar, PSR J0538+2817, with a transverse velocity of 357\,km\,s$^{-1}$. Both these velocities are 
directed away from the center of the supernova 
remnant. This leads to  kinematic age of 30\,000\,yr, when both stars originated in the center of the supernova remnant. The high-velocity of the B0V star indicates that the pre-disruption system 
had 
a relatively short orbital period (less than 15\,d), because only the disruption of a close system can have produced such a high velocity. The short orbital period then means that prior to the 
supernova there must have been extensive mass transfer in the system, and that at the time of the explosion the exploding star was the less massive star of the system. The fact that the system still 
was disrupted means that the disruption can only be due to the high kick velocity imparted to the neutron star at its birth. This is direct proof of birth kicks of neutron stars. Also, it is direct 
proof of the \citet{Blaauw1961} mechanism for producing runaway early-type stars.
Finally, also the fourth prediction was confirmed, by the discovery of the double radio pulsar PSR J0737-3039, in which the recycled (old) pulsar A with a pulse period of 22.7\,ms and a weak 
magnetic 
field, has as companion J0737-3039B, which is a normal “garden variety” young pulsar with a period of 2.773\,s and a strong magnetic field ($B \approx 5\times10^{11}$\,G) at the 
age around $10^7$\,yr \citep{Lyne2004}. Clearly, this is the second-born neutron star in the system, whose birth event induced the orbital eccentricity of the system.
The fact that the orbital eccentricity of this system (and of about half of the double neutron stars) is quite low ( $\leq{0.20}$) indicates that, like in the BeXBs with nearly circular orbits, the 
second-born  neutron stars in these systems received hardly any kick velocity at birth. As mentioned above, this means that these neutron stars resulted either from electron-capture collapse, or from 
the collapses of ultra-stripped iron cores.

\section {From Black-hole HMXBs to WRXBs and double black holes} 

\noindent
 {\underline{\it The Black-hole X-ray binaries}}
 
The bulk of the about 60 known BH-XBs \citep{CorralS2016} consists of Low-Mass X-ray Binaries (LMXBs), with typical donor masses $\leq 2{M_{\rm\odot}}$. These are the 
so-called “Soft X-ray Transients” or “X-ray Novae”, which may be dormant for decades, and then go into a bright X-ray outburst. A spectacular example is the outburst of V404 Cygni (GS2023+338) in 
2016. Although the black holes in these systems no doubt are remnants of massive stars, these systems are not HMXBs, and therefore I will not discuss them here.
We know only five black-hole HMXBs, of which Cygnus X-1 is the best-known example. Table~\ref{tab1} lists these 5 systems. One of them is the recently discovered BeXB MWC 656 
\citep[][Ribo, this volume]{Casares2014}. Two of the BH-XBs are in the Large Magellanic Cloud and one of them is in M33. The only system that probably is massive enough to produce a double 
black hole 
is the latter one: M33 X-7, a $15.7{M_{\rm\odot}}$ black hole with a $70{M_{\rm\odot}}$ companion in a 3.45-day orbit \citep{Orosz2007}. Its evolutionary origin has been described 
by \citet{Valsecchi2010}.
The rareness of BH-HMXBs with respect to the BH-LMXBs is, of course, a selection effect: the donor stars in the LMXB systems are very long-lived ($\geq 10^{9}$\,yr), while those in the HMXB 
systems 
are very short-lived ($\leq 5\times10^{6}$\,yr). The chances for observing a BH-LMXB, even though these turn on only occasionally, is therefore much larger than the chances of finding a BH-HMXB.
The most massive BH-HMXBs are the ones which we expect to evolve into WRXBs, which we will consider now.

\begin{table}
  \begin{center}
  \caption{The five known Black Hole High-Mass X-ray Binaries}
  \label{tab1}
 {\scriptsize
  \begin{tabular}{|l|c|c|c|c|}\hline 
{\bf Name} & {\bf ${P_{\rm {orb}}}$ (d)} & {\bf ${M_{\rm {don}}}$}&{\bf${M_{\rm {BH}}}$}& {\bf {Ref.}} \\ 
   & & {\bf ${M_{\rm\odot}}$}&  {\bf ${M_{\rm\odot}}$} &  \\ \hline
Cyg X-1 & $5.6$ & $19.2(\pm{1.9})$ & $14.8(\pm{1.0})$ & $(1)$ \\ 

LMC X-1 & $3.9$ &  $31.8(\pm{3.5})$ & $10.9(\pm{1.4})$& $(2)$\\ 
LMC X-3& $1.7$ & $3.6(\pm{0.6})$ & $7.0(\pm{0.6})$ &$(3)$  \\ 
 MCW 656& $\sim {60}$ & $\sim{13}$ & $4.7(\pm{0.9})$ & $(4)$ \\ 
 M33 X-7&$3.45$ & $70(\pm{7})$ & $15.7(\pm{1.5})$ & $(5)$ \\ \hline

  \end{tabular}
  }
 \end{center}
\vspace{1mm}
 \scriptsize{
 {\it References:}
  $(1)$ \citet{Orosz2011}, $(2)$ \citet{Orosz2009} 
  $(3)$ \citet{Orosz2014}, $(4)$ \citet{Casares2014}, $(5)$ \citet{Orosz2007}}
\end{table}

\noindent
{\underline{\it The WR X-ray Binaries}}

The first-discovered system of this type is Cygnus X-3 (see section 4). This is one of the most spectacular X-ray binaries known, and the only X-ray binary to which once an entire issue of Nature 
Physical Science was devoted (vol. 239, October 23, 1972). On September 23, 1972 its radio brightness increased by a factor of over $10^{3}$ \citep{HjellmingB1972}, which 
was the start of several giant  radio outbursts, making it temporarily the brightest radio source in the sky. The source is right in the galactic plane, and the three hydrogen 21-cm radio absorption 
lines at different doppler shifts visible during its radio outbursts showed that the source is behind three spiral arms, yielding a distance of about 10 kpc. The evolution of its radio spectrum 
during the outbursts was exactly as observed in quasar outbursts, indicating that it was synchrotron emission of an expanding cloud of relativistic electrons with magnetic fields. The adiabatic 
expansion of this cloud produces the characteristic quasar-like evolution of its radio spectrum \citep{HjellmingB1972}. So, Cyg X-3 in 1972 was already a 
{\it{micro-quasar}} long before this name was introduced by \citet{mirabel1992}. Around this time also its 4.8\,h X-ray orbital period was discovered by Bert Brinkman 
\citep{Parsignault1972}, and following its radio outbursts it was discovered that it is a strong infra-red (IR) source with the same periodicity \citep{Becklin1972}. Because its radio 
behaviour and strong IR emission are totally different from what is observed for LMXBs, while its position right in the galactic plane strongly suggests that it 
is a Population I object, and its orbital period fitted exactly with the outcome of our calculations at that time of the later evolution of a HMXB, we suggested Cyg X-3 to be a close helium star plus 
compact star binary \citep{HeuvelLoore1973}. This indeed was later proven to be correct (see section 4). During radio outbursts its IR spectrum is that of a WN7 
star, during radio quiescence it is that of a WN5 star \citep{Hanson2000}. Its IR luminosity of $3\times 10^{39}$\,erg\,s$^{-1}$, requires a helium star with a mass in the range $8 
-12\,M_\odot$ \citep[e.g.][]{Crowther2007}.

\begin{table}
  \begin{center}
  \caption{The Wolf-Rayet X-ray Binaries. Except for Cygnus X-3 the data are taken from the compilation by \citet{Esposito2015}, where the original references to the data of the different 
systems 
can be found. The mass of the WR star in Cyg X-3 is estimated from its IR luminosity, as mentioned in the text. As explained in the text, the mass estimates of the compact stars in WRXBs are very 
uncertain, but on the basis of binary evolution, these compact stars are expected to be black holes.   }
  \label{tab2}
 {\scriptsize
  \begin{tabular}{|l|c|c|c|c|}\hline 
{\bf Galaxy} & {\bf Source} & {\bf Orbital Period} & {\bf WR mass}            & {\bf accretor mass } \\ 
                    &                     & {\bf ($h$)}               & {\bf ${M_{\rm\odot}} $} & {\bf ${M_{\rm\odot}}$} \\ \hline
IC 10           & X-1          & $34.9$                     & ${35}$                         & ${33}$(?) \\
NGC 300    & X-1                                     & $32.8$                    & $26$                            &${20}$(?) \\ 
NGC 4490  & CXOUJ123030.3+413853 & $6.4$                     & $-$                               & $-$ \\
NGC 253   & CXOUJ004732.0-251722.1 & $14.5$                 & $-$                               & $-$\\
Circinus    & CG X-1                                 & $7.2$                   & $-$                               & $-$ \\
M 101       & ULX-1                                   & $196.8$               &$19$                      & ${20}$(?) \\ 
Milky Way & Cyg X-3 &$4.8$ & $8-12$ & $\geq{3}$(?) \\ \hline

  \end{tabular}
  }
 \end{center}
\vspace{1mm}  
 \scriptsize{}
\end{table}

For a long time, Cyg X-3 was the only WRXB known, but in the past decade a half dozen such systems have been discovered in external galaxies, as summarized in Table~2. See for details 
\citet{Esposito2015} and the papers of Carpano and of Soria in this volume.

The orbital periods of all but one of the systems are of order one day or shorter, showing that the systems must be the result of drastic spiral in evolution. The masses of the WR stars in IC 10 X-1 
and NGC 300 X-1 were derived from the optical brightness of these stars. The masses of their compact companions are very uncertain, as discussed by \citet{Laycock2015} and 
Carpano and Soria (this volume). The reason is that these masses were derived from the observed radial velocity curves of the emission lines of the WR stars, which curves are almost 90 degrees out of 
phase with the radial velocity curves expected from the X-ray light curves of these binaries. Therefore, these radial velocity curves cannot be those of the centers of mass of the WR stars, but 
probably are due to shock-features in the WR wind.   

\noindent
{\underline{\it Formation of WRXBs, reason why their compact stars likely are black holes}}

There are basically three ways in which a close WRXB can be formed: (i) by CE evolution of a wide BH-HMXB; (ii) by stable Roche-lobe overflow of a BH-HMXB with a relatively short orbital period, and 
(iii) through homogeneous evolution of a massive binary with a very short orbital period. We consider now each of these mechanisms separately.

(i)	{\it{Formation by CE evolution}}

This model is similar to that for the NS-HMXBs, leading to the formation of double neutron stars, as depicted in Fig.\,2, but scaled up to higher initial stellar masses. This is the model 
proposed by \citet{Bogomasov2014} and \citet{Belczynski2016} for the formation of close double black holes. In this case one must, like in Fig.\,\ref{fig2} , 
start from a wide binary system. CE evolution then makes the orbit of the system shrink by a large factor, leading to a system like Cyg X-3.

(ii) {\it{Formation through stable Roche-lobe overflow from a BH-HMXB with a blue supergiant donor star and a relatively normal (short) orbital period}}.
 
When the HMXB consists of a blue supergiant and a black hole, Roche-lobe overflow from the supergiant to the black hole does not need to become unstable, like in the case of a neutron star 
companion, 
and CE evolution can be avoided, as was shown by  \citet{vandenHeuveletal2017} and \citet{Pavlovski2017}. The conditions for stable Roche-lobe 
overflow are: (i) the donor star should have a radiative envelope \citep{KingTB2000}, and (ii) the mass ratio q of compact star and donor is q $\geq {0.3}$ \citep[for references to 
papers in which this condition was derived see][]{vandenHeuveletal2017}; 
\citet{Pavlovski2017} found that for very massive systems the 
latter condition can even be relaxed to q $\geq {0.2}$. 

A confirmation of the stability of mass transfer in the case of BH-HMXBs with $q \geq{ 0.3}$ is provided by the system of SS 433, a 13-day binary consisting of an A-giant donor star and a compact 
star which is surrounded by a huge and very luminous accretion disk, which completely dominates the light of the system \citep[e.g.][]{KingTB2000}. In this system Roche-lobe overflow 
is going on, in which the bulk of the transferred matter is ejected by the compact star in the form of the famous relativistic jets (with $v=0.265c$ and a mass-loss rate of 
$10^{-6} M_\odot$yr\,$^{-1}$), and in the form of a very strong wind ($10^{-4} M_\odot$\,yr$^{-1}$) from the huge accretion disk of the compact star \citep{Begelman2006}.
The mass transfer in this system is stable, as it has been going on already for thousands of years, without the system going into CE evolution \citep{KingTB2000}. This is 
demonstrated by the presence of the large W50 nebula which surrounds the system and the shape of which has been strongly influenced by the precessing relativistic beams together with the disk wind 
\citep[e.g][]{Begelman2006}). The mass ratio of the compact star and the donor in this system is indeed $\geq {0.3}$, as the estimated component masses are ${M_{\rm d}} = 
$12.1$ 
(\pm ${3.3}$){M_{\rm\odot}}$ and ${M_{\rm c}} = $4.3$ (\pm ${0.8}$){M_{\rm\odot}}$ \citep{HillwigGies2008}. The latter values may be underestimated, as pointed out recently by 
\citet{Cherepashchuk2018}, which authors do, however, agree with a mass ratio of the system $\geq{0.3}$, which implies, like for the mass estimates of 
\citet{HillwigGies2008} that the compact star in this system is indeed a black hole.

Supergiant BH-HMXBs typically have orbital periods $\leq{15}$\,d and are evolutionary products of normal WR+O-star 
close binaries, in which the core of the WR star has collapsed to a black hole. 
During 
their further evolution with stable Roche-lobe overflow, such systems spiral-in due to the SS433-like 
evolution in which the transferred matter is ejected from the surroundings of the compact star, 
carrying off the specific orbital angular momentum of the latter. Such systems terminate as WRXBs with 
orbital periods of the order of about one day \citep{vandenHeuveletal2017}. 
Since this type of evolution cannot occur for NS-HMXBs, this implies that in WRXBs formed in this way the 
compact stars always are black holes.

The same is true for the WRXBs formed through CE evolution, since the progenitor 
stars of the WR stars are more massive than $30{M_{\rm\odot}}$; with such a massive donor star the orbital period 
required with a $1.4{M_{\rm\odot}}$ neutron star to survive the CEE-phase is many years 
(much longer than for the case of a black hole companion, which typically has a mass $\geq 5{M_{\rm\odot}}$). 
Systems 
with such extremely long periods are expected to be very rare, such that the surviving WRXBs also 
in this case will practically exclusively harbor black holes. 

\begin{figure}[b]
\begin{center}
 \includegraphics[width=2.4in]{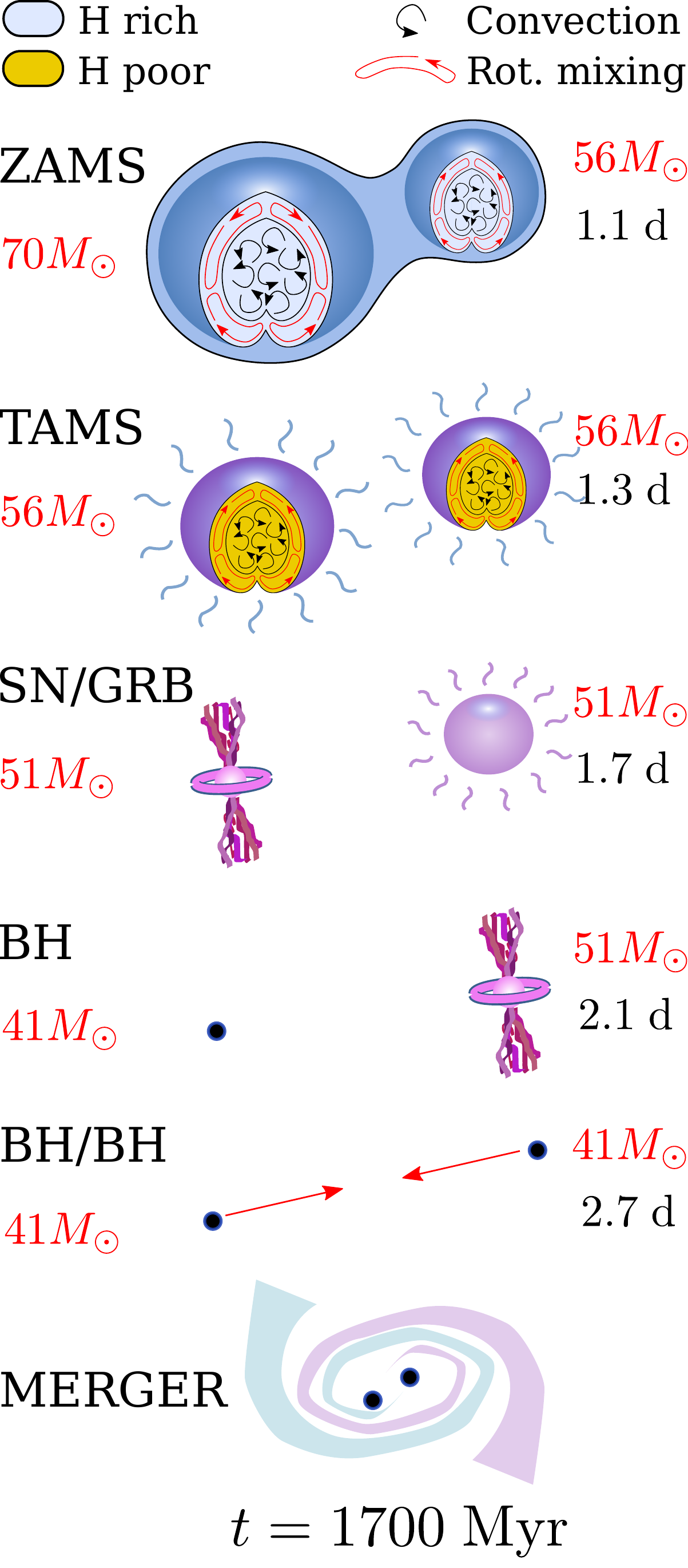}
 \caption{Homogeneous evolution of a very massive very close binary, orbital period {$ \leq{2 - 3}$\,d}, into a close double black hole, which merges within the lifetime of the universe. 
Explanation in the text. Figure courtesy Pablo Marchant.}.
   \label{fig3}
\end{center}
\end{figure}

\noindent
(iii)	{\it{Formation through homogeneous evolution}}

This model, for the formation of close double black holes, was proposed by \citet{Marchant2016} and \citet{MinkMandel2016}, based on the “homogeneous 
evolution” model of close very massive binaries with mass ratio close to unity,  put forward by \citet{Mink2008}. This model is based on the fact that in massive binaries 
with orbital periods $\leq {2-3}$\,d (such as are known in the Doradus region of the LMC) the strong tidal friction will cause the rotation period of the nearly equal-mass components to always be 
fully synchronized with the orbital period. This means that the two stars are kept in very rapid rotation. In such rapidly rotating massive stars, strong meridional circulation will develop, which 
keep the stellar material mixed throughout the star \citep{Maeder1987}. Thus, the helium produced by the hydrogen burning in the core will be fully mixed through the star, and the star will keep a 
fully homogeneous 
composition throughout its entire hydrogen-burning phase, and will end as a pure helium star: a WR star. This implies that, contrary to the case of normal massive stars, the stellar radius never 
increases during its evolution, and the star never overflows its Roche lobe. The final helium star is smaller than the original H-rich star. As the two stars will never have a mass ratio exactly 
equal 
to unity, the more massive component will be the first to become a WR star, while its companion then is still burning hydrogen, and will look like an O-star. By the time the WR star collapses to a 
black hole, its lower-mass companion may now have itself become a WR star, such that for a while the system will be a short-period WRXB. After the core collapse of the second WR star the system 
terminates as a close double black hole. This evolutionary sequence for forming double black holes is depicted in Fig.\,\ref{fig3} \citep[after][]{Marchant2017}.

\section{Summary and conclusions}

We have seen that:

(1)	The close double neutron stars are the later evolutionary products of wide neutron-star BeXBs, 
with orbital periods upwards of about one year. Neutron-star HMXBs with shorter orbital 
periods will merge and are expected to produce Thorne-Zytkow stars. Although this result has been 
known for over 40 years, and Thorne Zytkow stars should be quite common, so far never such an object 
has been identified with certainty.

(2)	Close double black holes (and black-hole neutron star systems) that formed through binary evolution, 
are later evolutionary products of the short-period WRXBs.

For the formation the latter systems three models have been identified: (i) through CE evolution from 
wide BH-HMXBs, which is basically the same model as that for the formation of double neutron 
stars, scaled up to higher masses; (ii) by in-spiral due to stable Roche-lobe overflow from BH-HMXBs 
with “normal” supergiant HMXB orbital periods, upwards from a few days. Only systems with orbital 
periods less than about 10 days will probably be able to terminate as double black hole systems with 
orbital periods short enough to merge within a Hubble time; (iii) by homogeneous evolution of 
massive close binaries with orbital periods $\leq {2-3}$ days and mass ratios $\geq {0.7-0.8}$.  

{\it{Acknowledgements}}:

I thank  Lida Oskinova and the SOC of this symposium for inviting me to present this overview. 
I thank Thomas Tauris, Ilya Mandel, Pablo Marchant, Selma de Mink and Chris Belczynski for enlightening 
discussions, which have much increased my understanding of the formation of double black holes. 
I thank Ron Taam and Natasha Ivanova for discussions on Common Envelope Evolution during more 
than 35 and more than 10 years, respectively. I thank Thomas Tauris for providing figures 2 and 3.

\end{document}